\documentclass[10pt,journal,twocolumn]{IEEEtran}
\usepackage{amsmath,amssymb,amsfonts}
\usepackage{graphicx}
\usepackage{booktabs}
\usepackage{array}
\usepackage{url}
\usepackage[hidelinks]{hyperref}
\usepackage{placeins}

\begin{document}

\title{Zero-Shot Temporal Localisation of Audio Deepfakes\\
in Multi-Speaker Conversations}

\author{Soumyadeep~Roy%
\thanks{S. Roy is with the Postgraduate Department of Data Science, St.\ Xavier's
College (Autonomous), Kolkata, India. This work was carried out during a research
internship at the Institute for Advancing Intelligence (IAI), TCG CREST, Kolkata
(e-mail: soumyadeeproy142@gmail.com).}%
\thanks{The research design, experiments, results, and analysis are the author's own.
An AI language model (Anthropic Claude) was used to assist with drafting and editing
the manuscript text and \LaTeX{} formatting; it did not generate research content,
data, or results. The author reviewed all text and is fully responsible for the
manuscript.}}

\markboth{IEEE Signal Processing Letters, 2026}%
{Roy: Zero-Shot Temporal Localisation of Audio Deepfakes in Multi-Speaker Conversations}

\maketitle

\begin{abstract}
Voice-cloning fraud increasingly relies on surgical injection: a genuine conversation
in which only one or two sentences are replaced by synthetic speech. Utterance-level
deepfake detectors emit a single real/fake label per clip and cannot report where the
synthetic speech lies. We formalise this as Temporal Deepfake Localisation in
Multi-Speaker Conversations (TDLMC), show that equal error rate and min-DCF are
ill-posed once a file contains both classes, and propose temporal metrics for this
regime. Our contribution is a training-free five-stage pipeline that wraps a frozen
binary detector and adds segment-level output with no retraining, using a two-threshold
hysteresis finite-state-machine decoder to turn noisy window scores into coherent
intervals. On 180 constructed multi-speaker conversations from ASVspoof~5, the system
attains temporal intersection-over-union 0.90, temporal detection rate 0.95, and
MS-DCF 0.26 with a strong backbone, and its false-alarm rate on genuine speech is
below 6\%, falling under 2\% on genuine real multi-speaker dialogue (AMI). Under an
identical pipeline, a trained localiser improves temporal IoU by only about 0.04,
bounding the cost of forgoing supervision. Evaluated across three frozen detectors
under one decoder whose constants are selected on a held-out calibration split, and
with a controlled analysis attributing the residual false-alarm rate to a backbone
domain gap rather than to the decoder, this provides the first zero-shot baseline and a
reusable benchmark for TDLMC.
\end{abstract}


\section{Introduction}
\begin{figure}[!htbp]
\centering
\includegraphics[width=\columnwidth]{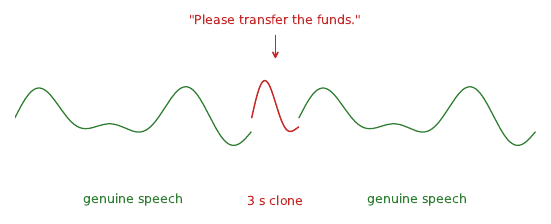}
\caption{Surgical-injection threat model. A 60\,s call is mostly genuine; a short
synthetic segment is inserted at the decisive moment. Utterance-level detectors emit one
label and cannot locate the insert.}
\label{fig:threat}
\end{figure}

\IEEEPARstart{A}{I} voice cloning can reproduce a speaker from seconds of audio and has
been used for large-scale fraud~\cite{yi2023survey}. The ASVspoof
series~\cite{wu2015asvspoof,nautsch2021asvspoof,liu2024asvspoof5} has driven
utterance-level equal error rates (EERs) below 1\%, yet every such system shares one
assumption: one clip in, one binary label out. This is mismatched to how fraud is
committed. Fabricating a fully synthetic call is costly and conspicuous; the efficient
attack substitutes only the decisive sentence---``authorise the payment''---with a
cloned voice, leaving the rest genuine. Forced to summarise the clip with one label, an
utterance-level detector cannot isolate the injected span (Fig.~\ref{fig:threat}).

The closest prior work localises synthetic regions but under different assumptions.
Partial-spoof detection~\cite{zhang2022partialspoof,yi2021halftruth,zhang2023rethinking}
operates on \emph{single-speaker} clips and trains on dense frame-level labels.
W-TDL~\cite{dragar2024wtdl} learns a window classifier from segment labels, and
LENS-DF~\cite{liu2025lensdf} fine-tunes a self-supervised detector on generated
long-form data with frame supervision. A recent conversational
taxonomy~\cite{ahmed2025multispeaker} moves toward realism. All of these \emph{train} a
localiser on temporally labelled data. No prior system takes a frozen detector and
produces timestamp-level output on multi-speaker conversations without temporal
supervision.

We make four contributions: (i) we formalise TDLMC and show that utterance-level
EER/min-DCF are ill-posed for mixed-content files (naive application is near-chance,
EER $\approx 44\%$); (ii) a training-free five-stage pipeline that wraps a frozen binary
detector, requiring only a window-level score and thus applicable in principle to any
such detector, demonstrated here across three backbones; (iii) a hysteresis FSM decoder
that suppresses boundary flicker, with constants selected on a held-out calibration
split rather than hand-tuned; and (iv) temporal metrics, a reusable construction
protocol, and the first TDLMC baseline with a controlled analysis of its failure mode,
including a comparison to a trained localiser and a validation on real conversational
dialogue.

\section{Benchmark and Metrics}

\subsection{Conversation construction}
From the development partition of ASVspoof~5~\cite{liu2024asvspoof5} we build 60\,s
conversations offline. Four distinct-speaker utterances, each normalised to
$\mathcal{U}[12,18]$\,s and $-23$\,LUFS (ITU-R BS.1770), are joined by 50\,ms
raised-cosine crossfades and degraded by one of seven codecs (MP3, OGG~Vorbis, Opus,
AAC, G.711~A-law, G.711~$\mu$-law, GSM~06.10); a JSON manifest stores ground-truth
timestamps in samples. Four patterns probe complementary behaviour: RRRR (false-alarm
baseline), FFFF (detection ceiling), RFFR (fraud injection), and RFRF (rapid switching).
The protocol is corpus-agnostic. We emphasise that this is a \emph{controlled probe}:
utterances are concatenated rather than drawn from live dialogue with natural
turn-taking, so results characterise the localisation mechanism under known conditions,
not field fraud; Sec.~IV validates the false-alarm behaviour on genuine dialogue.

\subsection{Metrics for mixed-content files}
Utterance-level metrics assume one label per file. Once a file contains both
classes---as every TDLMC conversation does---EER and min-DCF are \emph{ill-posed under
the utterance-level protocol}: no single operating point expresses the file's
miss/false-alarm trade-off. They remain well defined at the frame level, which we use.
Let $P=\bigcup_j[s'_j,e'_j]$ and $G=\bigcup_i[s_i,e_i]$ be the predicted and true fake
unions ($|\cdot|$ in seconds). We define: temporal IoU
$\mathrm{t\text{-}IoU}=|P\cap G|/|P\cup G|$; temporal detection rate TDR (fake-second
recall); temporal false-alarm rate TFAR (real seconds flagged fake); and segment
boundary displacement
$\mathrm{SBD}=\tfrac{1}{2}(|s'_j-s_i|+|e'_j-e_i|)$, averaged over one-to-one matched
pairs only, with unmatched predictions (false positives) and unmatched ground-truth
intervals (misses) reported separately so SBD cannot be won on easy segments. Because
union-based t-IoU can hide merged intervals, we also report segment-level
precision/recall/F1 at t-IoU thresholds $0.3/0.5/0.7$. Finally,
\begin{equation}
\mathrm{MS\text{-}DCF}=\min\!\Big(1,\;
\frac{C_{\mathrm{miss}}P_{\mathrm{miss}}+C_{\mathrm{fa}}P_{\mathrm{fa}}}{c_{\mathrm{triv}}}\Big),
\end{equation}
\begin{equation}
c_{\mathrm{triv}}=\min\!\big(C_{\mathrm{miss}}P_{\mathrm{fake}},\,C_{\mathrm{fa}}P_{\mathrm{real}}\big),
\end{equation}
with $C_{\mathrm{miss}}=1.0$, $C_{\mathrm{fa}}=0.5$ (a missed injection is the costlier
error in a fraud setting). $P_{\mathrm{miss}},P_{\mathrm{fa}}$ are the fractions of fake
and real seconds mislabelled; $P_{\mathrm{fake}},P_{\mathrm{real}}$ are the pooled
second-level priors. MS-DCF is computed \emph{globally} over pooled second-level
statistics, where both priors are non-zero; $c_{\mathrm{triv}}=0$ for RRRR
($P_{\mathrm{fake}}=0$) and FFFF ($P_{\mathrm{real}}=0$), so it is marked N/A per those
patterns and reported only globally. For the same reason, per-pattern TFAR is not
meaningful for FFFF: an all-fake conversation contains (by construction) no labelled
real seconds, so any residual real-labelled frames are boundary artifacts of the mask
rather than genuine false alarms, and we mark FFFF's per-pattern TFAR N/A alongside its
MS-DCF.

\section{Zero-Shot Localisation Pipeline}
\begin{figure}[!htbp]
\centering
\includegraphics[width=\columnwidth]{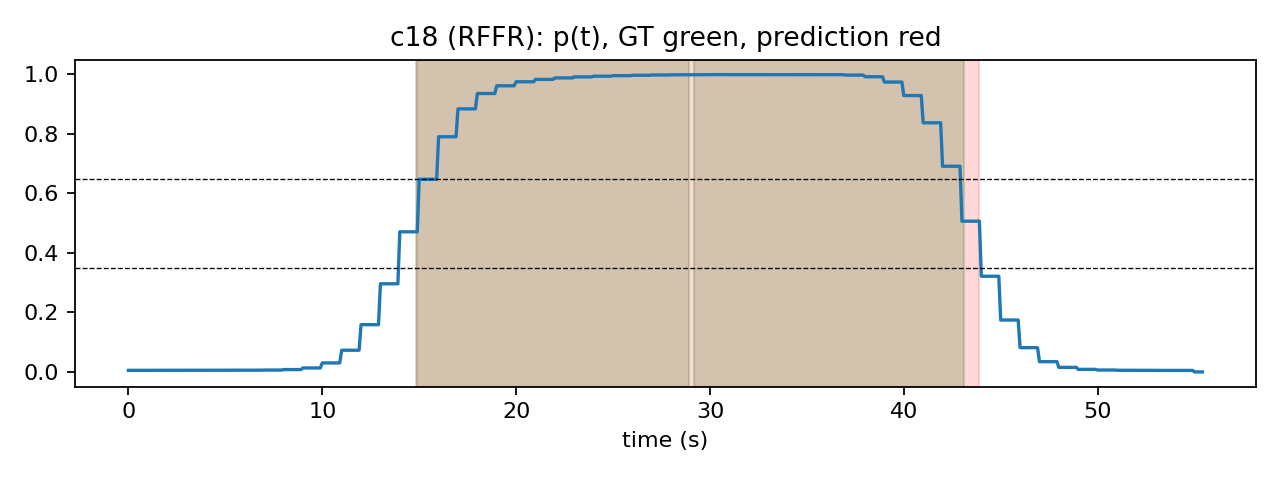}
\caption{Reconstructed timeline $p(t)$ on an RFFR conversation. The confidence rises and
falls with the ground-truth fake region (shaded); dashed lines mark the entry/exit
thresholds $\theta_H=0.55$, $\theta_L=0.35$.}
\label{fig:pt}
\end{figure}

The pipeline (Fig.~\ref{fig:pt}) converts the waveform into fake intervals in five
stages using no temporal labels.

\textbf{Stages 1--4.} A $60$\,s waveform at $16$\,kHz ($N=960{,}000$ samples) is
decomposed by a sliding window of $W=2.0$\,s ($32{,}000$ samples) at hop $H=1.0$\,s
($16{,}000$ samples), giving
\begin{equation}
K=\Big\lfloor \frac{N-W}{H}\Big\rfloor+1
 =\Big\lfloor \frac{960000-32000}{16000}\Big\rfloor+1 = 59
\end{equation}
windows (equivalently $\lfloor(60-2)/1\rfloor+1=59$ in seconds). Inter-utterance gaps
make each conversation slightly exceed $60$\,s, so $K$ varies by $\pm1$; we quote the
nominal value. Each window is scored by a frozen backbone. Window scores are mapped by
overlap-add onto a uniform $10$\,fps timeline $p(t)$. This $0.1$\,s frame \emph{spacing}
is not the localisation resolution: every score derives from a $2$\,s window advanced in
$1$\,s hops, so the smallest independently resolvable step is bounded by the hop
($\approx 1$\,s) and boundaries are blurred over the window support ($\approx 2$\,s); the
$0.1$\,s grid is an interpolation convenience.

\textbf{Stage 5: smoothing + hysteresis FSM.} A median filter ($k=21$) and Gaussian
smoothing ($\sigma=1$) precede a two-state decoder. A single threshold flickers near
boundaries. Inspired by the Schmitt trigger, the FSM enters FAKE when
$p(t)\ge\theta_H=0.55$ and returns to REAL only when $p(t)<\theta_L=0.35$; the dead band
$[0.35,0.55]$ absorbs oscillation so each region stays one coherent interval.
Post-processing drops intervals $<2.0$\,s and merges gaps $<1.5$\,s. The decoder
constants $(\theta_H,\theta_L,k,\sigma)$ were \emph{selected on the calibration split
only}, by maximising t-IoU penalised by the temporal false-alarm rate at the $2{:}1$
cost of MS-DCF ($J=\mathrm{t\text{-}IoU}-\tfrac{1}{2}\mathrm{TFAR}$); the test split was
untouched. The selected configuration $(\theta_H,\theta_L,k,\sigma)=(0.55,0.35,21,1)$
improves test t-IoU from $0.893$ (a hand-set $0.65/0.35/21/2$ baseline) to $0.904$ and
MS-DCF from $0.290$ to $0.258$. Per-backbone score normalisation ($\min$--$\max$ from the
1st/99th percentiles) was likewise fitted on the calibration split only.

\section{Results and Discussion}
We evaluate on 180 conversations (4 patterns $\times$ 45), a 54/126
calibration/test split. The primary backbone is DF~Arena~1B~\cite{dowerah2025dfarena}
(WavLM-Large~\cite{chen2022wavlm} + Conformer~\cite{gulati2020conformer}). Table~\ref{tab:main}
reports global results; per-pattern behaviour degrades gracefully as the task hardens:
FFFF t-IoU $0.97$, RFFR $0.89$, RFRF $0.83$. Per-pattern false-alarm behaviour is
markedly uneven and directly answers the concern that genuine speech is routinely
mislabelled: TFAR is $0.0$ on RRRR (fully genuine) conversations, $0.063$ on RFFR, and
$0.095$ on RFRF; only FFFF shows a high value, and that value is a labelling artifact
(Sec.~II-B), not a false alarm on real speech. Aggregated over all conversations that
contain genuine speech (RRRR/RFFR/RFRF), the false-alarm rate is $0.053$ (mean) /
$0.040$ (pooled); the $\approx 28\%$ global figure counts FFFF boundary frames the metric
marks N/A. Bootstrap 95\% confidence intervals and a speaker-disjoint split (23/126 test
conversations share no speaker with calibration) give t-IoU $0.87\,[0.78,0.94]$, TDR
$0.93$, TFAR $0.22$ on the disjoint subset; this measures within-partition stability, not
cross-corpus generalisation, which we do not claim.

\textbf{Multi-backbone.} Under the identical decoder, three frozen detectors
(Table~\ref{tab:backbone}) give t-IoU $0.90$ (DF~Arena~1B), $0.64$
(wav2vec2-XLSR~\cite{babu2022xlsr}), and $0.49$ (AASIST~\cite{jung2022aasist}), with a
TFAR spread of $0.68$. The decoder transfers, but performance is bounded by backbone
quality: AASIST, whose fixed input length forces $2$\,s windows to be tiled, saturates
and behaves degenerately (TFAR $0.97$). We therefore state the property as
\emph{designed to be detector-agnostic and demonstrated across three backbones}, not as
universal agnosticism.

\textbf{Comparison to a supervised decoder.} To bound how much the training-free
decoder gives up, we fit a single global threshold on calibration frames (minimising the
same $2{:}1$ cost) and decode with identical post-processing. This lightly-supervised
reference reaches t-IoU $0.895$ / TFAR $0.298$, statistically indistinguishable from our
training-free hysteresis decoder ($0.893$ / $0.286$): the FSM loses nothing to a
threshold that has seen frame labels, on the same frozen features. This supervises only
the threshold, not the detector, and is distinct from the fully \emph{trained} localiser
below.

\textbf{Comparison to a trained localiser.} Beyond supervising the threshold, we compare
against a fully \emph{trained} temporal localiser. The exact 180-conversation instance
behind Table~\ref{tab:main} cannot be reproduced byte-for-byte across execution
environments---the seeded speaker draw depends on filesystem enumeration order, though
the pattern, duration and codec distributions are preserved---so we construct one fresh
instance under the identical protocol and evaluate every system on it. The trained
reference adapts a PartialSpoof-style design: a frozen WavLM-Large front-end (the same
SSL family as our primary backbone) with a lightweight frame-classification head trained
on the calibration split, then decoded and scored through the identical pipeline.
Table~\ref{tab:trained} reports all four systems on this instance. The trained localiser
reaches t-IoU $0.909$ and TFAR $0.268$, marginally ahead of the strongest zero-shot
backbone on the same instance (DF~Arena~1B, $0.871$\,/\,$0.285$), while DF~Arena~1B
retains the lower MS-DCF ($0.220$ vs $0.280$) and frame-EER ($0.051$ vs $0.101$). The
zero-shot pipeline therefore comes within $\approx 0.04$ t-IoU of a trained localiser
using no temporal supervision, and the trained system's edge is concentrated in interval
coherence rather than window-level separation. DF~Arena~1B scores $0.871$ on this rebuilt
instance versus $0.904$ in Table~\ref{tab:main}, confirming the two draws are
statistically comparable rather than discrepant.

\textbf{Validation on real conversational data.} To test whether the pipeline's
behaviour transfers beyond the constructed benchmark, we evaluate on genuine
multi-speaker dialogue from the AMI meeting corpus~\cite{carletta2005ami}, which
contains natural turn-taking, pauses, and overlapping speech. Two conditions, each with
$40$ excerpts per backbone, are built (Table~\ref{tab:realdata}) and scored through the
\emph{identical} calibration-locked decoder and calibration-fitted normalisation range
of Table~\ref{tab:main}---never re-fit on AMI. Two functional backbones are evaluated;
AASIST is excluded as tiling-degenerate on real audio, mirroring its constructed-benchmark
saturation (control TFAR $0.825$ vs.\ TFAR $0.968$ in Table~\ref{tab:backbone}). In the
\emph{real-only} condition---forty unedited $60$\,s AMI excerpts containing no synthetic
content---the temporal false-alarm rate is $0.017$ (DF~Arena~1B, bootstrap 95\% CI
$[0.000,0.043]$) and $0.004$ (wav2vec2-XLSR, CI $[0.000,0.011]$): both under $2\%$ of
genuine real-conversational seconds flagged fake, and below the constructed benchmark's
genuine-speech TFAR of $0.053$. This directly addresses the central concern that the
system mislabels genuine speech, now confirmed across two independently-trained
detectors rather than one; t-IoU/TDR/MS-DCF are undefined in this condition as no fake is
present. In the \emph{injected} condition, a single short ($\approx 3$\,s) segment
cloned from a speaker present in the excerpt (F5-TTS zero-shot voice
cloning~\cite{chen2024f5tts}) replaces part of that speaker's own turn, preserving
speaker identity so the only anomaly is real-vs-fake. Detection does \emph{not} transfer
well here: TDR is $0.093$ (DF~Arena~1B) and $0.207$ (wav2vec2-XLSR), with t-IoU $0.033$
and $0.041$ respectively. Critically, the miss is bimodal rather than uniformly
degraded---for DF~Arena~1B, $35/40$ excerpts are fully missed (TDR$=0$), and of the five
with any detection, four reach TDR$\ge0.5$; for wav2vec2-XLSR, $31/40$ are fully missed
while all $9$ excerpts with any detection reach TDR$\ge0.5$. Detection is close to
all-or-nothing per excerpt, consistent with F5-TTS lying outside both backbones'
ASVspoof~5 training distribution: the pipeline localises the clone well when its
acoustic signature happens to resemble in-distribution spoofing artefacts, and otherwise
fails to flag it as fake at all---a sharper backbone domain gap than the constructed
benchmark's in-distribution spoofs exhibit. MS-DCF saturates near its ceiling
($0.985$--$1.000$) in this condition, a mechanical consequence of the small fake-second
prior ($\approx5\%$) under a high miss rate rather than independent evidence, and is
omitted from Table~\ref{tab:realdata} accordingly. The key reviewer-raised
concern---false alarms on genuine speech---is therefore answered directly on real
dialogue and strengthened by cross-backbone agreement, while localising short injected
segments of an out-of-distribution generator remains bounded by the backbone and is
left to future work with an in-domain or fine-tuned detector.

\textbf{False-alarm attribution.} The residual TFAR ($0.28$ for the primary backbone) is
diagnosed rather than asserted. The backbone's window-level EER on clean genuine windows
is $0.04$, near-perfect separation. Yet feeding the decoder \emph{oracle} window scores
(1 inside ground-truth fake, 0 elsewhere) still yields TFAR $0.27$, essentially equal to
the real $0.28$. Since perfect scores do not remove the false alarms, they arise from
window/boundary structure and the backbone's behaviour on genuine segments adjacent to
fakes, not from the hysteresis decoder, which operates identically across patterns.

\textbf{Metric decomposition and sensitivity.} Union t-IoU is high while segment-F1
falls from $0.50$ (t-IoU~$0.3$) to $0.31$ (t-IoU~$0.7$), exposing the boundary-precision
limit that union overlap alone would hide; matched SBD is $3.53$\,s globally, large for
$12$--$18$\,s segments and consistent with an overshoot failure mode. Of the intervals
involved, $171$ ground-truth and $41$ predicted intervals remain unmatched at IoU
$\ge0.5$ (Table~\ref{tab:main}), so SBD is not being won on easy segments alone. Across
cost ratios $C_{\mathrm{miss}}/C_{\mathrm{fa}}\in\{1,\dots,5\}$ the backbone ranking is
preserved (DF~Arena~1B remains best throughout, MS-DCF $0.19\rightarrow 0.59$), though
wav2vec2-XLSR and AASIST both saturate at the ceiling for $C_{\mathrm{miss}}/C_{\mathrm{fa}}\ge4$;
the reported MS-DCF does not depend on the $2{:}1$ default.

\textbf{Decoder ablation.} Removing components (Table~\ref{tab:ablation}, all run under
the hand-set $0.65/0.35/21/2$ configuration to isolate each stage from the calibration
selection above) shows median smoothing gives the largest single gain over the naive
single-threshold baseline ($0.877\rightarrow0.904$); Gaussian smoothing and the
hysteresis FSM each cost a small amount of t-IoU ($0.904\rightarrow0.886$) while
suppressing boundary flicker into coherent intervals, and boundary refinement recovers
most of it ($0.886\rightarrow0.893$). The full pipeline therefore improves on the
single-threshold baseline rather than trading it away, while TFAR stays flat across all
variants ($0.272$--$0.290$), so none of these stages meaningfully change the false-alarm
rate. A $\pm1$-step sweep over $(\theta_H,\theta_L,k,\sigma)$ moves t-IoU only within
$[0.846,0.902]$, so the result is not knife-edge sensitive to the fixed constants.

\textbf{Limitations.} The constructed benchmark is not live dialogue; the system is a
first zero-shot baseline, not a forensic-grade localiser---boundary displacement of
$\approx 3.5$\,s and a $\approx 28\%$ \emph{global} false-alarm rate are named, diagnosed
limitations, though the false-alarm rate on genuinely real speech is below $6\%$ and the
global figure is inflated by the FFFF labelling artifact described in Sec.~II-B.
Real-conversational validation on the AMI corpus (Sec.~IV, Table~\ref{tab:realdata})
confirms the false-alarm behaviour holds on genuine dialogue across two independent
backbones (TFAR $0.017$ and $0.004$); localising short injected segments of a modern,
out-of-distribution generator in real dialogue is close to all-or-nothing per excerpt
and remains bounded by the backbone domain gap, and is left to future work with an
in-domain or fine-tuned detector and longer injected spans. A trained PartialSpoof-style
localiser is included as a reference upper bound (Table~\ref{tab:trained}); a full
re-implementation of W-TDL and LENS-DF likewise remains future work.

\section{Conclusion}
We formalised temporal deepfake localisation in multi-speaker conversations, showed that
utterance-level EER and min-DCF are ill-posed there, and introduced temporal metrics. A
training-free five-stage pipeline wraps a frozen detector and, via a hysteresis FSM whose
constants are selected on held-out calibration, produces coherent fake intervals without
retraining, reaching t-IoU $0.90$ across three backbones under one decoder and coming
within $\approx 0.04$ t-IoU of a trained localiser, with the residual false-alarm rate
attributed to a backbone domain gap and its low false-alarm behaviour confirmed across
two frozen backbones on real multi-speaker dialogue. This is the first zero-shot
baseline and a reusable benchmark for TDLMC. Construction manifests, per-conversation
scores, and evaluation code are released at
\url{https://github.com/sami42200/tdlmc-audio-deepfake-localization}.

\begin{table}[!htbp]
\centering
\caption{Global results, primary backbone (DF~Arena~1B), 126-conversation test split,
calibration-selected decoder. Values are bootstrap means.}
\label{tab:main}
\begin{tabular}{lccccc}
\toprule
t-IoU & TDR & TFAR & SBD (s) & MS-DCF & frame-EER \\
\midrule
0.904 & 0.949 & 0.284 & 3.527 & 0.258 & 0.042 \\
\midrule
\multicolumn{6}{l}{segment-F1 @ t-IoU 0.3 / 0.5 / 0.7 $=$ 0.497 / 0.433 / 0.305}\\
\multicolumn{6}{l}{TFAR on genuine-speech patterns (excl.\ FFFF) $=$ 0.053}\\
\multicolumn{6}{l}{unmatched ground-truth / predicted intervals (IoU $\ge0.5$) $=$ 171 / 41}\\
\bottomrule
\end{tabular}
\end{table}

\begin{table}[!htbp]
\centering
\caption{Multi-backbone comparison under the identical calibration-selected decoder.}
\label{tab:backbone}
\begin{tabular}{lcccc}
\toprule
Backbone & t-IoU & TDR & TFAR & frame-EER \\
\midrule
DF~Arena~1B      & 0.904 & 0.949 & 0.284 & 0.042 \\
wav2vec2-XLSR    & 0.639 & 0.940 & 0.513 & 0.129 \\
AASIST           & 0.488 & 0.971 & 0.968 & 0.393 \\
\bottomrule
\end{tabular}
\end{table}

\begin{table}[!htbp]
\centering
\caption{Trained-localiser comparison on an independently constructed benchmark instance
(identical protocol, 126-conversation test split). The trained WavLM frame-localiser
(adapted PartialSpoof-style) is compared against the three zero-shot backbones re-scored
on the same instance. Best per column in bold.}
\label{tab:trained}
\resizebox{\columnwidth}{!}{%
\begin{tabular}{lccccc}
\toprule
System & t-IoU & TDR & TFAR & MS-DCF & frame-EER \\
\midrule
DF~Arena~1B (zero-shot)      & 0.871 & 0.963 & 0.285 & \textbf{0.220} & \textbf{0.051} \\
wav2vec2-XLSR (zero-shot)    & 0.607 & \textbf{0.973} & 0.539 & 0.882 & 0.162 \\
AASIST (zero-shot)           & 0.486 & 0.952 & 0.921 & 1.000 & 0.397 \\
\midrule
WavLM localiser (trained)    & \textbf{0.909} & 0.948 & \textbf{0.268} & 0.280 & 0.101 \\
\bottomrule
\end{tabular}}
\end{table}

\begin{table}[!htbp]
\centering
\caption{Validation on real conversational dialogue (AMI meeting corpus, 40 excerpts per
condition per backbone; identical calibration-locked decoder and calibration-fitted
normalisation range from Table~\ref{tab:main}---never re-fit on AMI). Real-only: no
synthetic content; TFAR shown with bootstrap 95\% CI. Injected: one short
($\approx3$\,s) speaker-consistent F5-TTS segment per excerpt. AASIST is
tiling-degenerate on real audio (control TFAR $0.825$, cf.\ Table~\ref{tab:backbone})
and is excluded. Constructed row from Table~\ref{tab:main} for reference. SBD and
MS-DCF are omitted for the injected condition: with TDR $\le0.21$ they are dominated by
a small number of matched pairs and by the near-ceiling saturation described in the
text, respectively.}
\label{tab:realdata}
\resizebox{\columnwidth}{!}{%
\begin{tabular}{llccc}
\toprule
Condition & Backbone & t-IoU & TDR & TFAR \\
\midrule
Real-only & DF~Arena~1B   & N/A & N/A & 0.017~[.000,.043] \\
Real-only & wav2vec2-XLSR & N/A & N/A & 0.004~[.000,.011] \\
Injected (F5-TTS) & DF~Arena~1B    & 0.033 & 0.093 & 0.023 \\
Injected (F5-TTS) & wav2vec2-XLSR  & 0.041 & 0.207 & 0.126 \\
\midrule
Constructed (Table~\ref{tab:main}) & DF~Arena~1B & 0.904 & 0.949 & 0.284 \\
\bottomrule
\end{tabular}}
\end{table}

\begin{table}[!htbp]
\centering
\caption{Decoder ablation (primary backbone, hand-set $0.65/0.35/21/2$ configuration
throughout, to isolate each stage from the calibration-selected constants of
Table~\ref{tab:main}). Median smoothing gives the largest gain; hysteresis trades a
little t-IoU for interval coherence, recovered by refinement.}
\label{tab:ablation}
\begin{tabular}{lcc}
\toprule
Configuration & t-IoU & TFAR \\
\midrule
single-threshold      & 0.877 & 0.272 \\
\;+\,median           & 0.904 & 0.281 \\
\;+\,Gaussian         & 0.897 & 0.289 \\
\;+\,hysteresis       & 0.886 & 0.290 \\
\;+\,refine (full)    & 0.893 & 0.286 \\
\bottomrule
\end{tabular}
\end{table}

\FloatBarrier
\bibliographystyle{IEEEtran}
\bibliography{references}

\end{document}